\documentstyle[editedvolume]{crckapb} 
\input epsf

\def\solar{\ifmmode_{\mathord\odot}\else$_{\mathord\odot}$\fi}
\def\deg{\ifmmode^\circ\else$^\circ$\fi\ }
\def\etal{et~al.\ }
\def\gtsima{$\; \buildrel > \over \sim \;$}

\begin{opening}
\title{GASDYNAMICS AND STARBURSTS
       IN INTERACTING GALAXIES}

\author{J. CHRISTOPHER MIHOS}
\institute{Department of Astronomy\\
           Case Western Reserve University}

\end{opening}

\runningtitle{GASDYNAMICS AND STARBURSTS IN INTERACTING GALAXIES}

\begin{document}


\begin{abstract}

The onset of gaseous inflows and central activity in interacting
galaxies is driven largely by induced bars in the host galaxies.
The stability of galaxies against growing bar modes is a direct 
function of their structural properties --- galaxies with central 
bulges or low disk surface densities are more stable against central 
starbursts than are bulgeless or disk-dominated systems. Low surface 
brightness galaxies prove less prone to bar formation and central 
starbursts than do normal high surface brightness galaxies. This 
stability of LSB disks also resolves many of the dynamical pitfalls 
encountered when attempting to link poststarburst ``E+A'' galaxies 
to interactions involving normal high surface brightness galaxy progenitors.

\end{abstract}

\section{Introduction}

Overwhelming evidence indicates that galactic collisions can lead to a 
large scale redistribution of gas in galaxies, driving strong nuclear 
inflows and fueling central activity (starburst and/or AGN) in many 
interacting systems. However, a one-to-one correlation between
interactions and central starbursts is not evident --- many interacting
systems show only modest star forming activity, distributed throughout
the body of the galaxy. What, then, determines the gasdynamical and
star forming response of a galaxy to a gravitational encounter?
Detailed N-body simulations of interacting systems have shown that the
onset of gaseous inflows is intimately tied to the formation of
global bars, which act to drive gas inwards to the central regions.
As such, the question of induced {\it star} formation becomes one of 
induced {\it bar} formation --- that is, the onset of inflow and activity 
is determined by a galaxy's stability against growing bar modes.

I describe how the structural properties of galaxies can influence
the gasdynamical and star-forming response of galaxies
to an interaction. I focus first on major mergers and the effects of
central bulges, then turn to more subtle ``flyby'' encounters and the
role of disk surface density in driving starburst activity. We find that
differences in galaxy structure lead to significantly different responses;
much of the variance in the properties of interacting systems can be
traced to differences in the progenitor galaxies.

\section{Gasdynamics in Major Mergers}

Major mergers of equal mass disk galaxies are thought to result in the most 
dramatic starburst events. The ``ultraluminous'' infrared galaxies (ULIRGs)
are prime examples of this process, where $\sim 10^{12}$ M\solar\ of gas has
been driven into their central regions, fueling intense 
(L$_{IR} > 10^{12}$ L\solar) activity (see, e.g., Sanders \& Mirabel 1996). 
To study the evolution of such
mergers, we employ $N$-body models to 
follow the combined gravitational, hydrodynamic, and star-forming evolution
of galaxies experiencing a merging event (Mihos \& Hernquist 1994ab; 1996).

We contrast models in which the merging galaxies have different structural
properties --- in particular, galaxy models with and without central bulges.
We employ a system of units wherein the disk mass M$_d = 1$, the
scale length of the disk $h=1$, and the gravitational constant $G=1$.
In both models, the galaxies consist of an exponential disk of stars and
gas ($M_{gas} = 0.1$) embedded in a spherical dark matter halo
with mass $M_h = 5.8$ and core radius $\gamma=1$, truncated
at $r=10h$.  In the model which includes a central bulge, the bulge possesses
a Hernquist (1990) profile, with mass $M_b = 1/3$ and scale
length $a=0.2$. Rotation curves for the different models are shown in
Figure 1ab. Star formation is included via a simple Schmidt law: 
SFR$\sim \rho_{gas}^{1.5}$ (see Mihos \& Hernquist 1994b).
The galaxies are placed on (initially)
parabolic orbits, with a Keplerian pericenter of $R_p=2.5$. One
disk is exactly prograde, the other is inclined by 71\deg to
the orbital plane. Figure 2 shows the inflow and star forming
properties of each model; images of the models can be found in
Mihos \& Hernquist (1994a, 1996).

\begin{figure}
\epsfbox{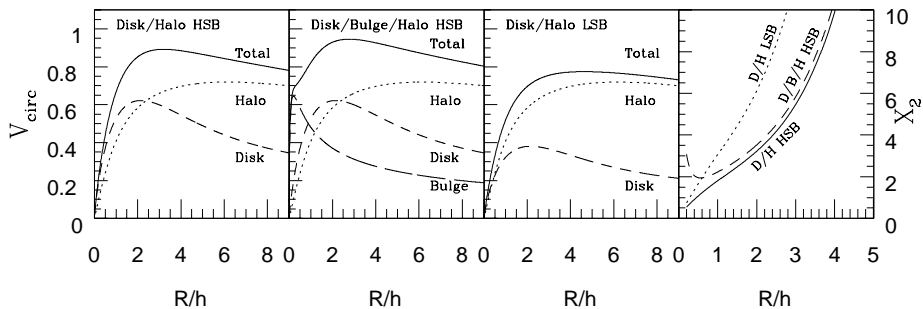}
\caption{Rotation curves of galaxy models. The first three panels show
the contribution of different components to the total rotation curve
of each galaxy, while the final panel shows the Toomre $X_2$ bar stability
parameter for each model.}
\end{figure}

Even though the interaction
parameters are identical, the star forming response of the two 
models is dramatically different.  The galaxies without bulges
rapidly develop strong bars --- the $m=2$ mode in the
stellar disk dominates the mass distribution shortly after the
galaxies first collide. Gas is compressed along this bar, forming
a gaseous bar which slightly leads the stellar bar.  This offset 
between the stellar and gaseous bars results in a net torque on
the gas, driving the strong inflow of gas into the nuclear regions. 
At this time, starburst activity is triggered in each nuclei while
the galaxies are still widely separated. These starbursts deplete 
the gas, so that when the galaxies ultimately merge, they are gas poor 
and lack the fuel to power any strong starburst associated with the 
final merging. As such, these models are poor representations of ULIRGs, 
which are gas-rich, late stage mergers with strong central activity. 
Evidently a major merger alone is not a sufficient condition to trigger 
ultraluminous activity; some other criteria is necessary.

\begin{figure}
\epsfbox{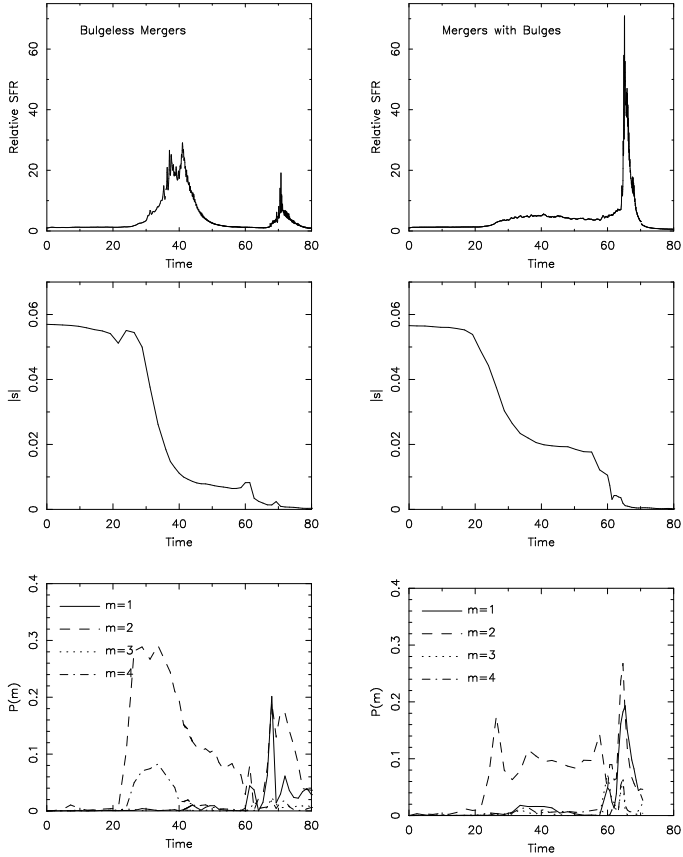}
\caption{Gas inflow and starburst activity in major mergers.
Left panels show the evolution of the bulgeless galaxy models,
while the right panels show the evolution of the models with
central bulges.
Top panels show the star formation history in the models (assuming
a Schmidt law for star formation); middle panels show the spin angular 
momentum of the inflowing gas in the prograde galaxy; and bottom panels 
show a Fourier decomposition of the stellar mass distribution in 
the prograde disk. The rotation period at the half mass radius is 
$T_{rot} \sim 15$. 
Initial collision occurs at T=24; the final merging occurs at T=65-70.}
\end{figure}

In contrast, the merger involving galaxies with bulges has a very 
different history of inflow and starburst activity. The presence
of a central bulge acts to stabilize the galactic disks against
the growth of bar instabilities; instead, the galaxies form 
tightly wound spiral arms which provide a weaker torque on the
disk gas. As a result, the gas inflow occurs in two stages.
Initially, the gas moves inwards, but ``hangs up'' at a radius
of a few kpc, where the bulge dominates the mass distribution 
and the disk torques are weaker. This weak inflow results in
only a modest enhancement of the star formation rate, and the
gas is not strongly depleted.  When the galaxies do finally
merge, the accompanying strong torques result in a second phase of
inflow --- the gas in both galaxies is very quickly driven into 
the center of the merger, and an extreme starburst event is triggered. 
Unlike the bulgeless merger, this merger with bulges has properties 
(morphology, gas content, starburst strength) which compare favorably 
with observed ultraluminous infrared galaxies. It is the internal 
dynamics of the merging galaxies which is the necessary criterion
for the formation of ULIRGs.

As these models demonstrate, the detailed response of galaxies to a 
merger depends critically on their stability against
the onset of global bar modes. This stability has been characterized
by the Toomre $X_2$ parameter (Toomre 1981): $X_2=\kappa^2R/4\pi 
G\Sigma_{disk}$. If $X_2<1$ (for a linearly rising rotation curve)
or $X_2<3$ (for a flat rotation curve), disks are susceptible to
growing $m=2$ modes. Fig 1d shows $X_2$ for the different models ---
by changing the shape of the rotation curve, the bulge acts to
stabilize the inner disk against bar modes. Without bulges, colliding disk 
galaxies become bar unstable on a dynamical timescale, and experience 
early inflows and central activity. Adding a bulge inhibits bar formation
and the associated early inflow, resulting in more dramatic activity when the
galaxies ultimately merge. Clearly these two model represent ``endpoints''
of a distribution of bulge:disk ratios in galaxies; the response 
of individual systems will depend in detail on the their structural 
properties and progenitor type.

\section{Flyby Encounters and LSBs}

The previous merger models show the connection between inflow, starbursts,
and disk stability. However, there is another path to disk stability besides
central bulges, and that is through lowered disk surface density.
If the density of the disk is sufficiently low (at fixed rotation
velocity), perturbations
cannot be amplified into strong bar modes, and bar-induced inflows are
suppressed. Such may be the case in low surface brightness (LSB) disk
galaxies, which have low disk surface densities and large dark matter
contents (de Blok \& McGaugh 1996, 1997). 

To examine how disk surface density influences inflow and star forming
activity during galaxy interactions, we look at the evolution of galaxies
experiencing an equal-mass, non-merging ``flyby'' encounters. Again,
two models are contrasted. The first, representing a high surface
brightness (HSB) disk galaxy, is the bulgeless disk/halo galaxy model
employed in the previous merger simulations. The second model,
representing a LSB disk galaxy, is simply the HSB model with the disk
surface density reduced by a factor of 2.5 --- i.e., $\Delta \mu_0 \sim$
1 mag/arcsec$^2$ for
a similar (M/L)$_*$. The rotation curve for this model is shown
in Figure 1; the low disk surface density results in a greater stability 
against growing disk modes than in HSB disk galaxies (as shown by the 
higher value of $X_2$), save for the very central regions
where the disk still contributes an appreciable amount of the total 
mass density. Finally, compared to HSBs, LSBs generally have 
a higher gas-to-baryonic mass ratio and flatter gas mass profiles
(de Blok \etal 1996; McGaugh \& de Blok 1997);
this is reflected in the LSB  model which possesses a flatter gas mass 
profile with $M_{gas}/M_{disk}=1/3$ (see Figure 3).

\begin{figure}
\epsfbox{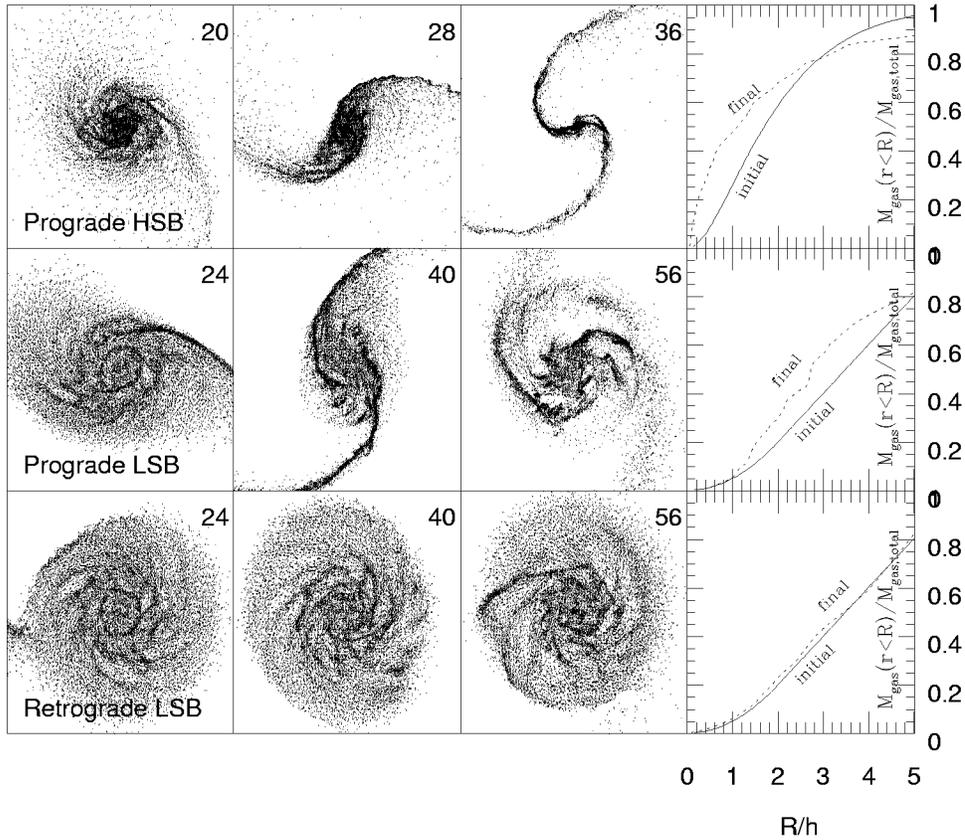}
\caption{The evolution of the ISM in flyby interactions. In each
strip, the first three panels show the morphology of the disk gas
after the galaxies pass closest approach (at T=24); the final panel
compares the initial and final (T=36 for the HSB disk, T=56 for the
LSB disks) cumulative gas mass profiles in each model. The rotation period 
at the half-mass radius for the disk is $T_{rot} \sim 15$. Note the
different timescales between the HSB and LSB models; the HSB model
quickly develops a bar which drives rapid inflow. The LSB models
evolve much more slowly.}
\end{figure}

The flyby interactions involve parabolic orbits with pericenter separation of
$R_p=10h$. Figure 3 shows the evolution of the ISM component in the different
models. In the prograde HSB encounter, the galaxy quickly develops a strong 
bar (see Mihos \etal 1997); gas is compressed along this bar and is rapidly 
driven inwards. By T=36, only one half-mass rotation period for the disk, 
already $\sim$ 30\% of the gas has been driven into the inner kpc (assuming a 
Milky Way scaling for the model); the calculation was stopped here, but 
inflow continues along the strong bar in the model. 

By contrast, the prograde LSB disk lacks sufficient self-gravity in the
disk to amplify the perturbation of the interaction into a strong bar.
Instead, the galaxy develops a milder oval distortion with strong
spiral arms. Gas is compressed along these arms, and fragments into
small clumps throughout the disk; presumably these would be sites of 
enhanced star formation. There is some mild inwards migration of gas
in the system, but the mass distribution in the inner scale length is
largely unchanged --- without a strong bar, little inflow into the
central regions occurs. We emphasize that the prograde geometry is the 
``worst case'' scenario for driving instabilities in disks, due to the
resonance between orbital and rotational motion; other geometries will
be much less damaging.
For example, the last set of panels in Figure 3 shows a retrograde LSB 
encounter --- very little evolution is observed, even during this relatively
close encounter.

Clearly the low disk surface density and high dark matter content of LSB
disks affords them a great deal of stability against bars and induced gas 
inflows. This stability actually strengthens with declining surface brightness
(as surface density decreases and dark matter becomes ever more dominant;
de Blok \& McGaugh 1997),
such that very low surface brightness galaxies may in fact be very stable
systems, not easily destroyed or turned into starburst HII galaxies by
casual interactions. Instead, mild evolution in 
surface brightness may result due to enhanced disk star formation. 

Finally, we emphasize that the stability described here pertains to
the {\it internal} amplification of perturbations into bar modes. This
discussion is most apropos to mild interactions or secular evolution of 
LSBs. Mergers or continual interactions (such as in a cluster environment) 
may act to overwhelm the disk stability and drive additional activity
even in these dark matter dominated systems.

\section{The Progenitors of E+A's}

``E+A'' galaxies are systems with spectral characteristics identifying 
them as having experienced a strong starburst in the past $10^9$ years, 
after which star formation ceased entirely. The fact that such systems 
are observed in both the cluster and field environments (Zabludoff \etal 
1996) argues in favor of formation mechanisms which are not cluster specific,
such as galaxy interactions and mergers (e.g., Lavery \etal 1992). While this picture of
mergers driving the formation of strong E+A systems works on a qualitative
basis, upon closer scrutiny several inconsistencies appear when trying
to invoke ``normal'' (i.e., HSB) spiral galaxies as the merging progenitors.
Alternatively, I argue that LSB progenitors may solve a number of these 
inconsistencies. To wit:

\begin{enumerate}

\item {\it High starburst mass fractions:} Extreme E+A galaxies may
have starburst mass fractions $M_{burst}/M_{*}$ \gtsima 0.2 (Couch \& Sharples
1987; Barger \etal 1996; Liu \& Green 1996). Such
high burst masses indicate the galaxies must have been {\it extremely}
gas-rich; Milky Way-type spirals simply lack sufficient gas to fuel
such a starburst. LSBs, on the other hand, are the most gas-rich objects
in the local universe (McGaugh \& de Blok 1997).

\item{\it Spatially extended burst populations:} In E+A's the 
young stellar population is often spatially extended, and not
confined to the nuclear region (e.g., Franx 1993; Caldwell \etal 1996). This 
is contrary to observations of local starburst galaxies, which are
predominantly nuclear starbursts (e.g., M82, or the ULIRGs).
However, the stability of LSB galaxies results in a 
{\it global} response to interactions, as gas fragments throughout the 
disk but is not driven into the galaxies' centers. Disk star formation,
rather than nuclear starbursts, is the likely outcome.

\item{\it E+A's in galaxy pairs:} The fact that some E+A's are found
in interacting pairs (Zabludoff 1996; Wirth 1996) raises a timescale problem --- if the interaction
caused the starburst, why is the system a {\it post}-starburst system,
even though the interaction is still ongoing? Some mechanism must
act to shut off star formation, independent of the interaction phase.
The low gas densities of LSBs, coupled with a threshold density
for star formation (e.g., Kennicutt 1989, van der Hulst \etal 1993), may provide such a shutoff 
mechanism. If the initial collision drives disk gas above threshold, 
strong disk star formation ensues. As this star formation depletes the gas, 
the gas density drops back below threshold, and star formation is 
stopped. Because the cessation of star formation is linked to {\it local} 
dynamical conditions, the starburst can terminate irrespective
of the dynamical phase of the interaction.

\item{\it Disky E+A's:} Some E+A galaxies are disky systems (Caldwell \etal
1996; Wirth 1996; Franx, this volume). Such E+A's cannot form through major
mergers, which destroy galactic disks. If interactions drive the 
formation of disky E+A's, they must involve low mass accretions or flyby 
passages. With starburst efficiencies lower in these types of interactions, 
the gas reservoir must be extremely large, as found in LSBs.

\end{enumerate}

Certainly not all E+A's need arise from interacting LSB disks -- many
E+A's show clear merger morphologies, or possess weaker burst strengths.
However, the interdependence described here between galactic structure, 
disk stability, gas inflows, and starbursts suggests that the variety of
progenitor galaxies available in the Universe necessarily dictates a
variety of poststarburst E+A galaxies. The E+A's which do not easily
fit into the picture of interactions involving normal ``Hubble-type''
spirals may in fact follow from LSB progenitors.

\end{document}